\documentclass{PoSi}
\usepackage{amsmath,amssymb}
\usepackage{upgreek}
\usepackage{xspace}

\def\eq#1{\begin{equation}#1\end{equation}}
\def\del{\partial}
\def\fref#1{Fig.\,\ref{#1}}

\def\funk{\textsc{Funk}\xspace}

\DeclareGraphicsExtensions{.pdf,.png,.jpg,.JPG}

\title{Search for hidden-photon dark matter with the \textsc{Funk} experiment}

\ShortTitle{Hidden-photon dark matter}

\author{D.~Veberi\v{c}$^a$,
A.~Andrianavalomahefa$^a$,
K.~Daumiller$^a$,
B.~D\"obrich$^b$,
\speaker{R.~Engel}$^a$,
J.~Jaeckel$^c$,
M.~Kowalski$^{de}$,
A.~Lindner$^d$,
H.-J.~Mathes$^a$,
J.~Redondo$^f$,
M.~Roth$^a$,
T.~Schwetz-Mangold$^a$,
C.M.~Sch\"afer$^a$,
R.~Ulrich$^a$
\newline
[The~\funk~Experiment]
\vspace{1mm}
\\
\llap{$^a$} Institute for Nuclear Physics, Karlsruhe Institute of Technology
(KIT), Karlsruhe, Germany\\
\llap{$^b$} Physics Department, CERN, Geneva, Switzerland\\
\llap{$^c$} Institute for Theoretical Physics, Heidelberg University, Germany\\
\llap{$^d$} Deutsches Elektronen Synchrotron (DESY), Zeuthen, Germany\\
\llap{$^e$} Department of Physics, Humboldt University, Berlin, Germany\\
\llap{$^f$} Department of Theoretical Physics, University of Zaragoza, Spain\\
\llap{$^*$} E-mail: \href{mailto:ralph.engel@kit.edu}{\rm ralph.engel@kit.edu}
}

\abstract{
\hfil\includegraphics[width=0.2\textwidth]{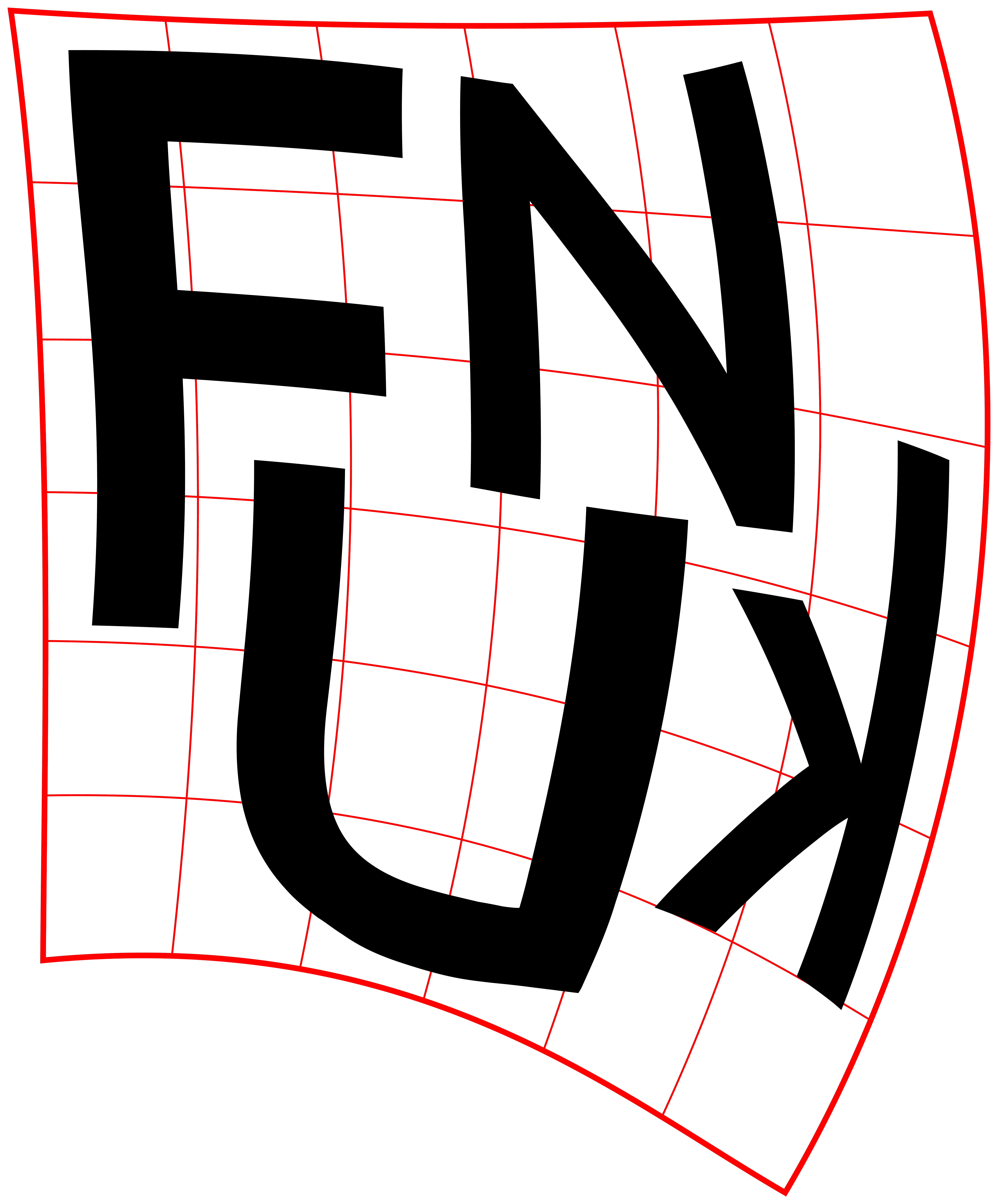}\hfil
\\
Many extensions of the Standard Model of particle physics predict a parallel
sector of a new U(1) symmetry, giving rise to hidden photons.  These hidden
photons are candidate particles for cold dark matter. They are expected to
kinetically mix with regular photons, which leads to a tiny oscillating
electric-field component accompanying dark matter particles. A conducting
surface can convert such dark matter particles into photons which are emitted
almost perpendicularly to the surface. The corresponding photon frequency
follows from the mass of the hidden photons. In this contribution we present a
preliminary result on a hidden photon search in the visible and near-UV
wavelength range that was done with a large, 14\,m$^2$ spherical metallic
mirror and discuss future dark matter searches in the eV and sub-eV range by
application of different detectors for electromagnetic radiation.
}

\FullConference{35th International Cosmic Ray Conference --- ICRC2017\\
                10--20 July, 2017\\
                Bexco, Busan, Korea}

\begin{document}

\section{Introduction}

\noindent
There is convincing astrophysical and cosmological evidence that a large
fraction of the energy density in the universe must be composed of invisible
non-baryonic matter or dark matter (DM)~\cite{Gelmini:2015zpa}. The most
explored options for explaining DM are extensions of the Standard Model (SM) of
particle physics predicting weakly-interacting massive particles (WIMPs). In
recent years attention has been turned also to alternative scenarios involving
weakly-interacting slim particles (WISPs) such as axions, axion-like particles,
or hidden photons (HP)~\cite{Nelson:2011sf}. WISPs could be non-thermally
produced in the early universe and survive as cold DM until today (see
e.g.~\cite{Jaeckel:2010ni,Ringwald:2012cu,Jaeckel:2013ija,Baker:2013zta} for
reviews).

HP are additional light U(1) gauge bosons that kinetically mix with the SM
photons~\cite{Holdom:1985ag}.  The effective Lagrangian can be written as
\eq{
\mathcal{L} =
  - \tfrac14(F_{\mu\nu}F^{\mu\nu} + X_{\mu\nu}X^{\mu\nu})
  + J^\mu A_\mu
  + \tfrac{m^2}2 X_\mu X^\mu
  - \tfrac\chi2 F_{\mu\nu}X^{\mu\nu},
}
where $F_{\mu\nu}=\del_\mu A_\nu-\del_\nu A_\mu$ is the field strength tensor
of the photon and $X_{\mu\nu}$ that of the corresponding HP field, and $J^\mu$
the charge current.  

Large phase space regions of the HP mass $m$ and mixing parameter $\chi$ are
compatible with the observed DM signatures~\cite{Arias:2012ly,Graham:2015rva}.
Many different methods are applied to search for hidden photons or to derive
constraints on their parameters. For example, laboratory experiments include
haloscopes, helioscopes, and light-shining-through-a-wall
methods~\cite{Arias:2012ly}.

In this paper we report about a search for HP as dominant DM particles with
\funk (Finding U(1)s of a Novel Kind).  \funk is an experiment that is based on
the recently proposed dish-antenna method \cite{Horns:2012jf,Jaeckel:2013sqa}.
A similar experiment, although with smaller sensitivity, has been performed in
Tokyo~\cite{Suzuki:2015sza,Suzuki:2015vka}.

\begin{figure}[t]
\def\figh{0.3}
\centering
\includegraphics[height=\figh\textwidth]{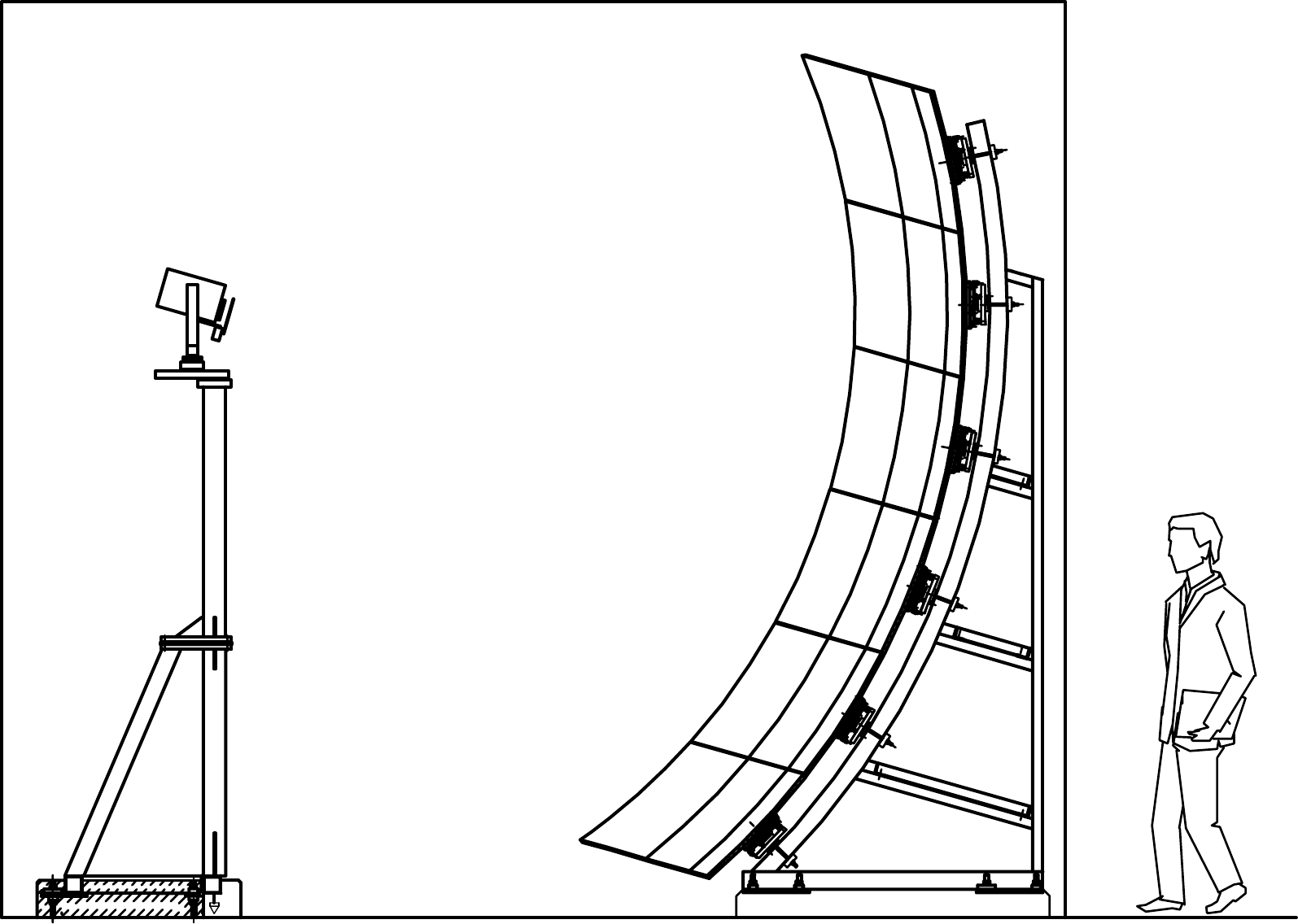}\hspace*{1cm}
\includegraphics[height=\figh\textwidth]{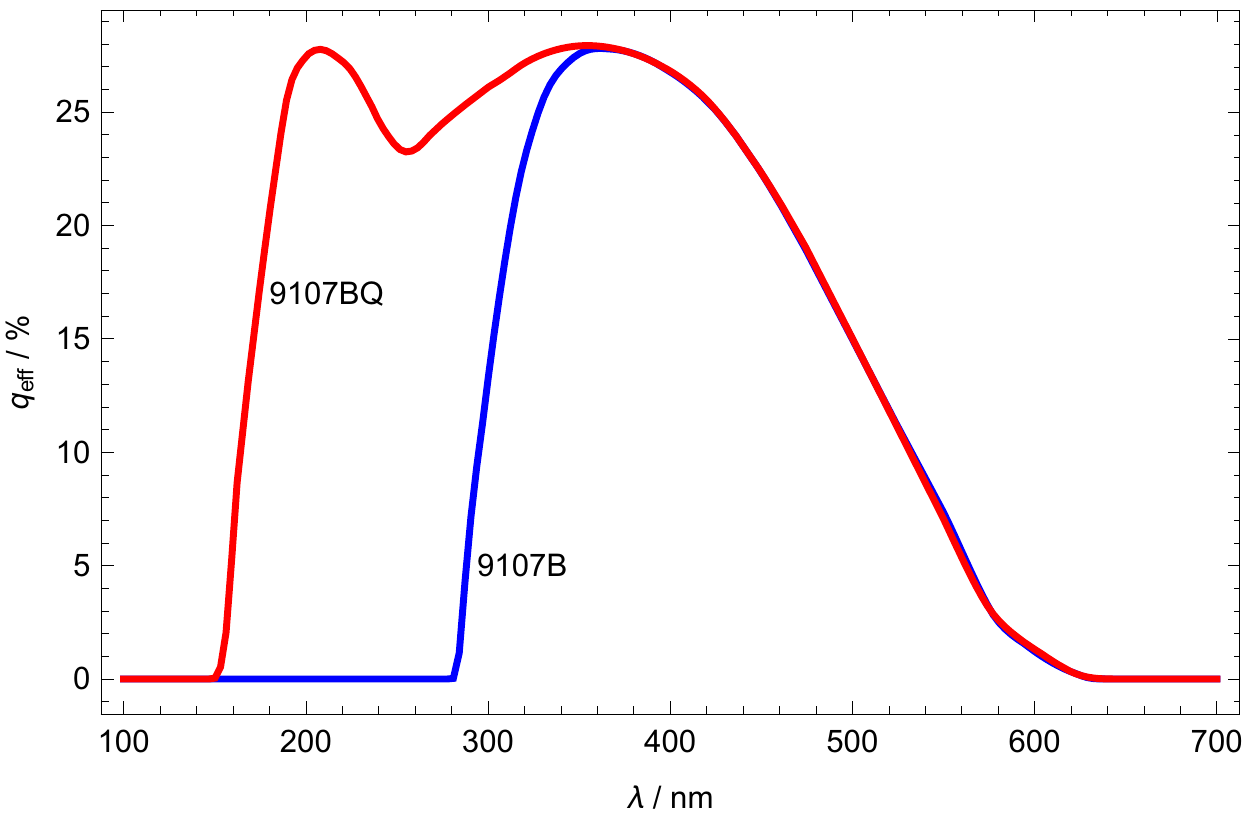}
\caption{\emph{Left:} Schematic view of the \funk experiment with a pillar
supporting the PMT camera in the center of the spherical mirror consisting of
$6\times 6$ segments. The box outlines a $495{\times}431{\times}435$\,cm$^3$
experimental volume. \emph{Right:} The quantum efficiency of the PMT used as
photon sensor~\cite{et-manual}.  The UV-extended sensitivity of the used PMT
9107BQ is shown in red.}
\label{f:mirror}
\end{figure}

\begin{figure}[t]
\def\figh{0.35}
\centering
\includegraphics[height=\figh\textwidth]{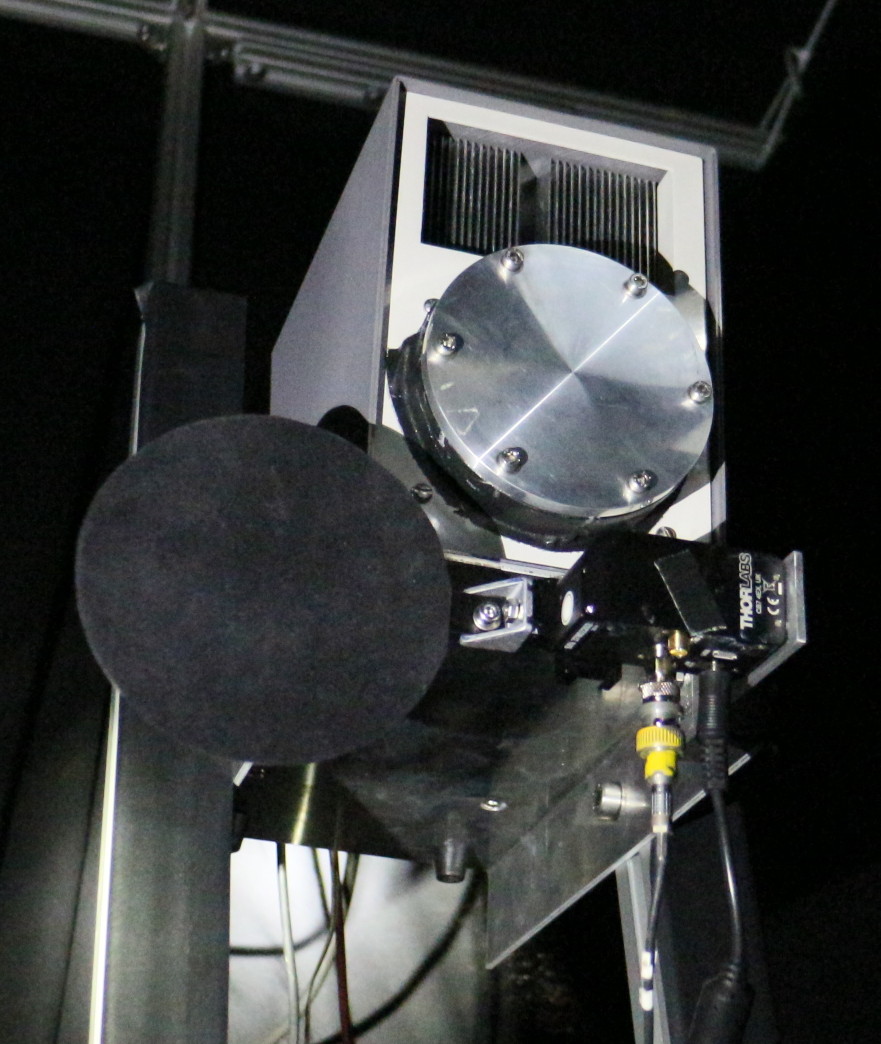}\hspace*{1cm}
\includegraphics[height=\figh\textwidth]{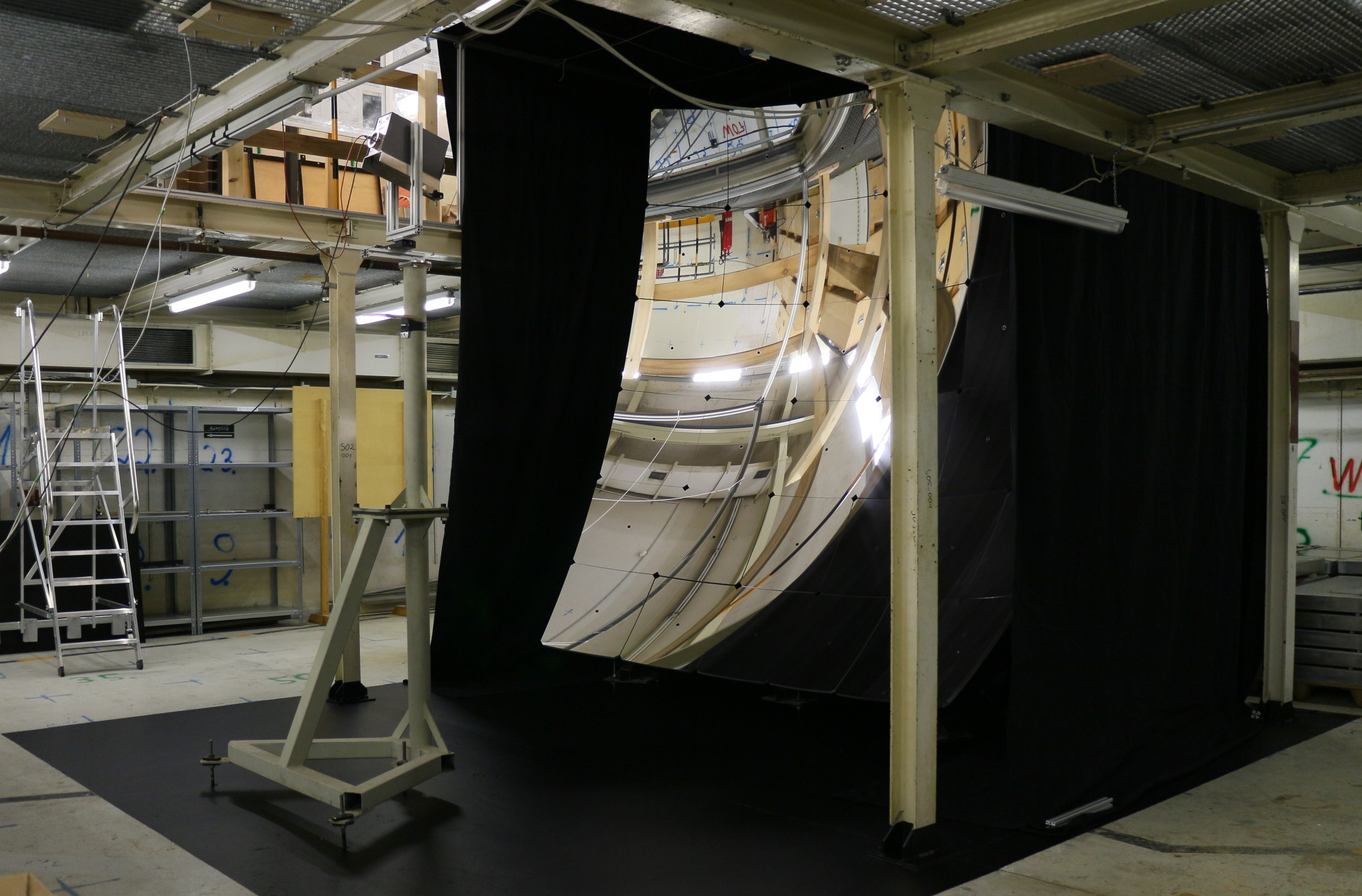}
\caption{\emph{Left:} A photograph of the PMT camera (closed with circular
aluminum lid) with a Thorlabs motorized flipper carrying an obscuring shutter
(black circle in front, in the ``open'' position).  \emph{Right:} The
photograph of the \funk experimental area, which is wrapped on top and the
sides with a double-layer cover (black plastic sheet over a cotton curtain).
The ground is treated with a mat-black paint.}
\label{f:setup}
\end{figure}

\begin{figure}[t]
\def\figh{0.4}
\centering
\includegraphics[height=\figh\textwidth]{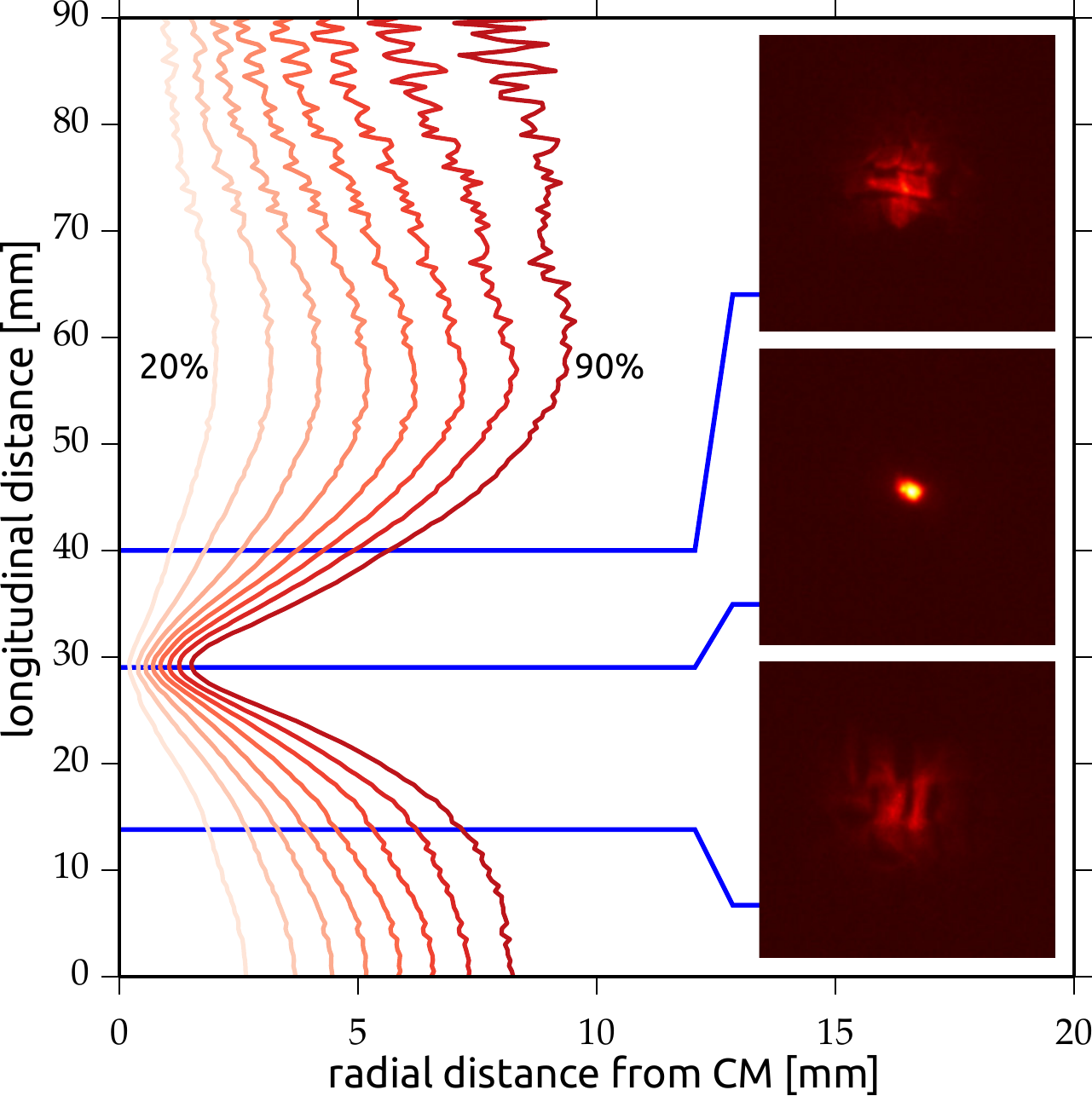}\hspace*{1cm}
\includegraphics[height=\figh\textwidth]{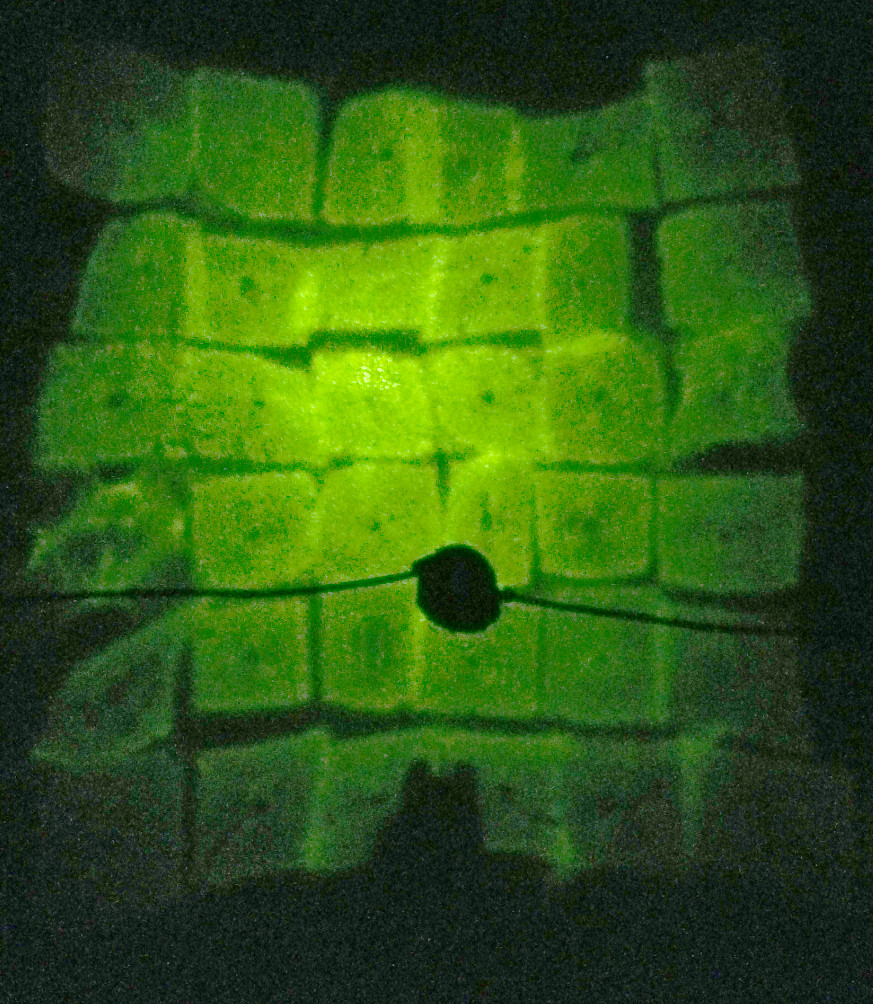}
\caption{\emph{Left:} Contours of radial distance from the image center of mass
(CM) containing certain amount of total light (from 20\% to 90\% in steps of
10\%) for the final, realigned assembly of the mirror. The three insets show
spot images in various positions marked by blue lines (middle line for the
smallest spread). 90\% of the light is contained within a spot of radius
$\lesssim2$\,mm. \emph{Right:} An image obtained when the LED is driven
off-center towards the mirror. Reflected beams are not converged yet so that
they form a matrix where each of the green squares corresponds to individual
mirror segments, revealing their aberrations. The central black dot is a shadow
of the LED with two horizontal power cables.}
\label{f:focus}
\end{figure}


\begin{figure}[t]
\def\figh{0.35}
\centering
\includegraphics[height=\figh\textwidth]{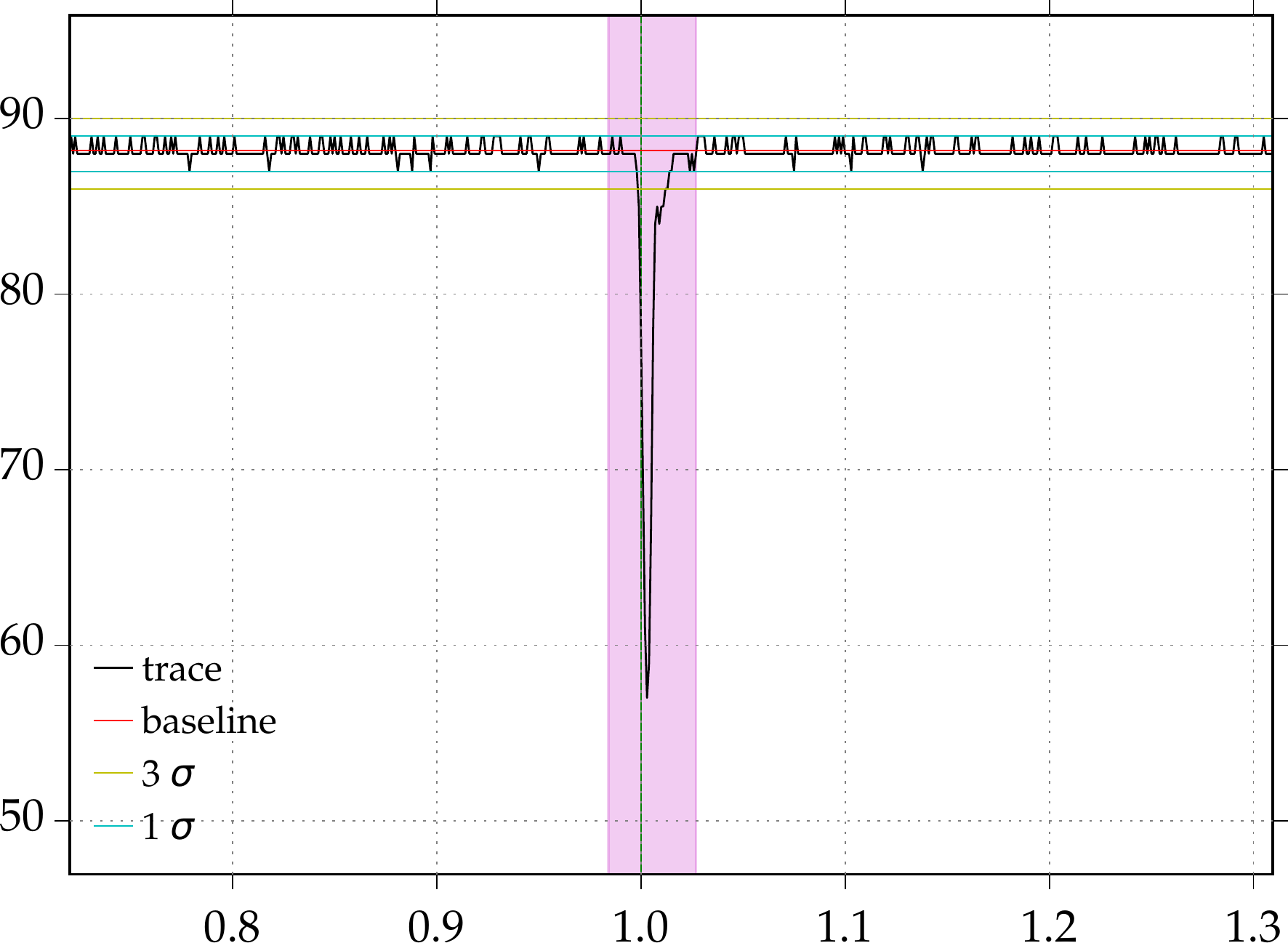}\hfil
\includegraphics[height=\figh\textwidth]{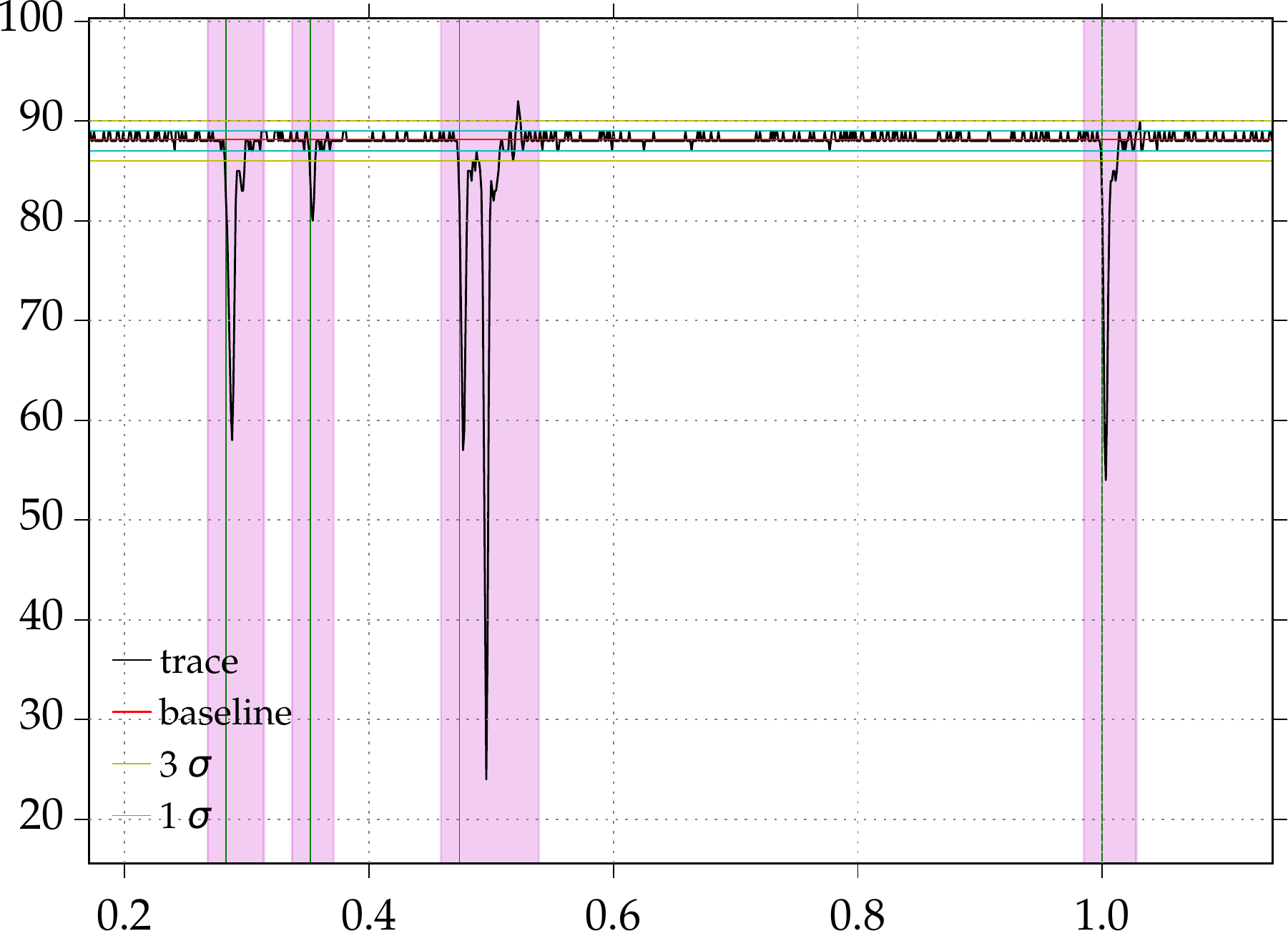}
\caption{\emph{Left:} An example of a typical single-photon trace.
\emph{Right:} An example of a trace with multiple pulses within the
$1.6\,\upmu$s trigger window which gets rejected by the cosmic-ray cut.}
\label{f:traces}
\end{figure}

\begin{figure}[t]
\def\figh{0.37}
\centering
\includegraphics[height=\figh\textwidth]{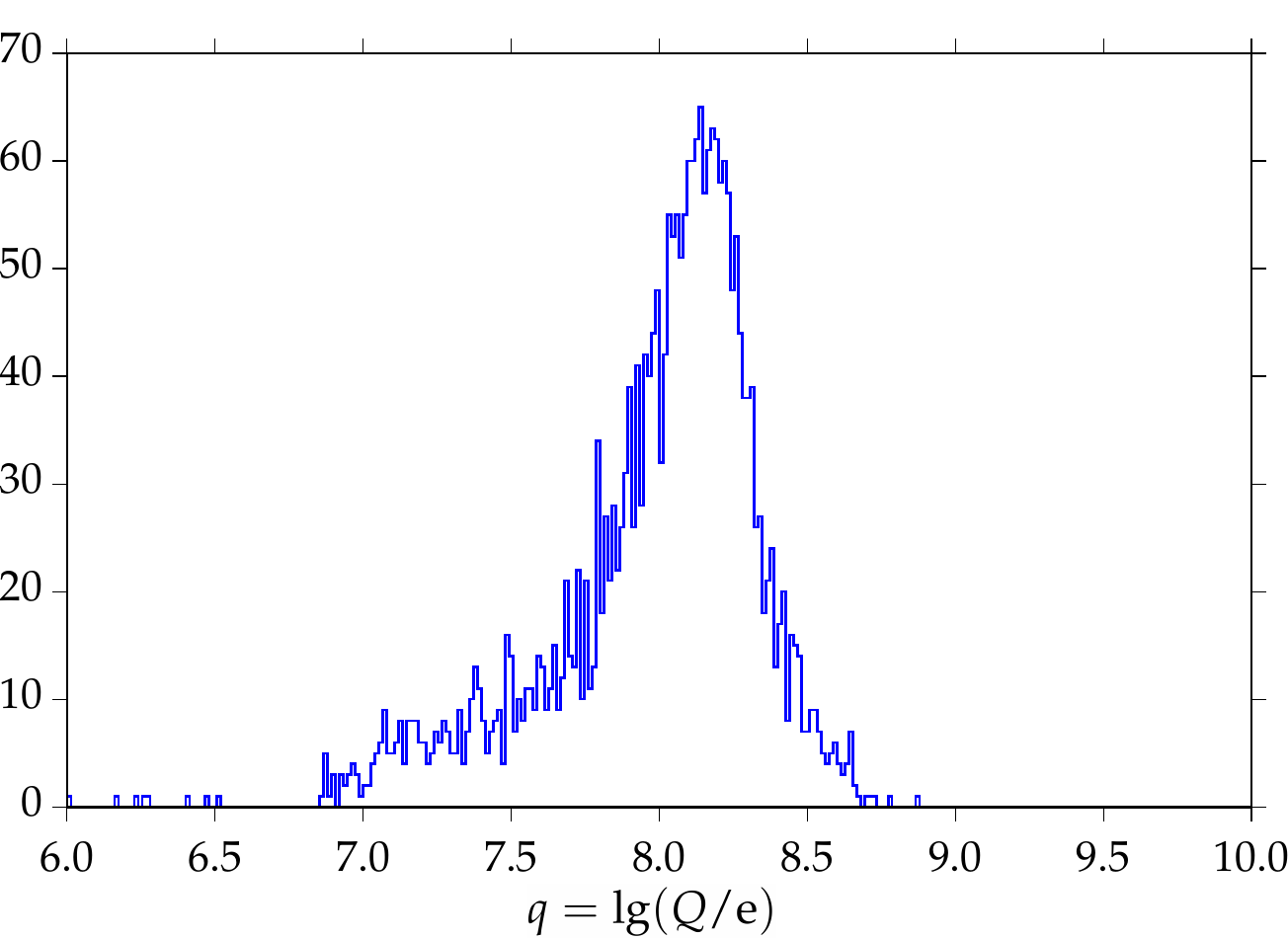}\hfil
\includegraphics[height=\figh\textwidth]{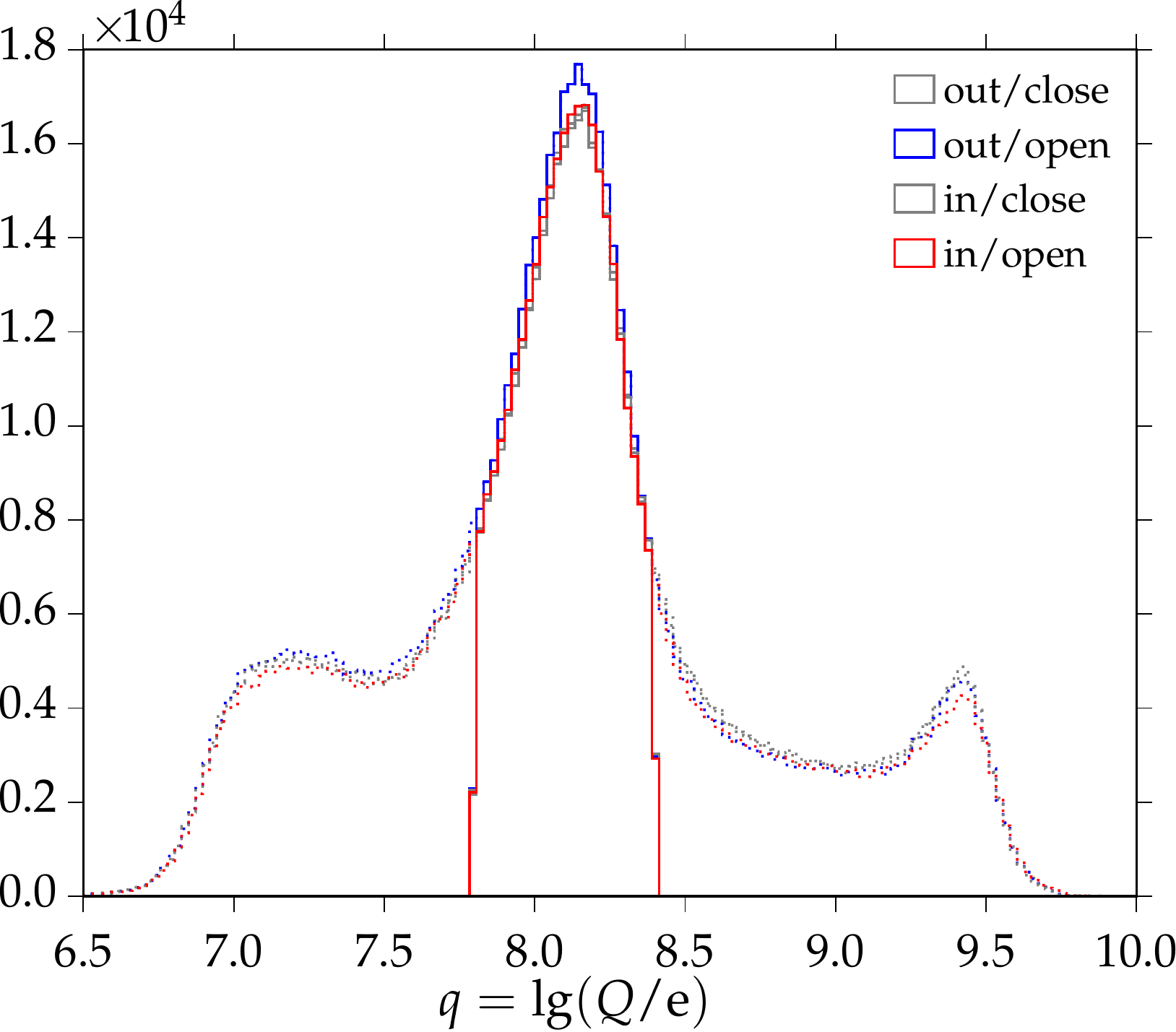}
\caption{\emph{Left:} Charge distribution for single photo-electron signals taken with an LED flasher;
\emph{Right:} Charge distribution of a \funk run (without external light source).}
\label{f:charge}
\end{figure}

\begin{figure}[t]
\centering
\includegraphics[height=0.4\textwidth]{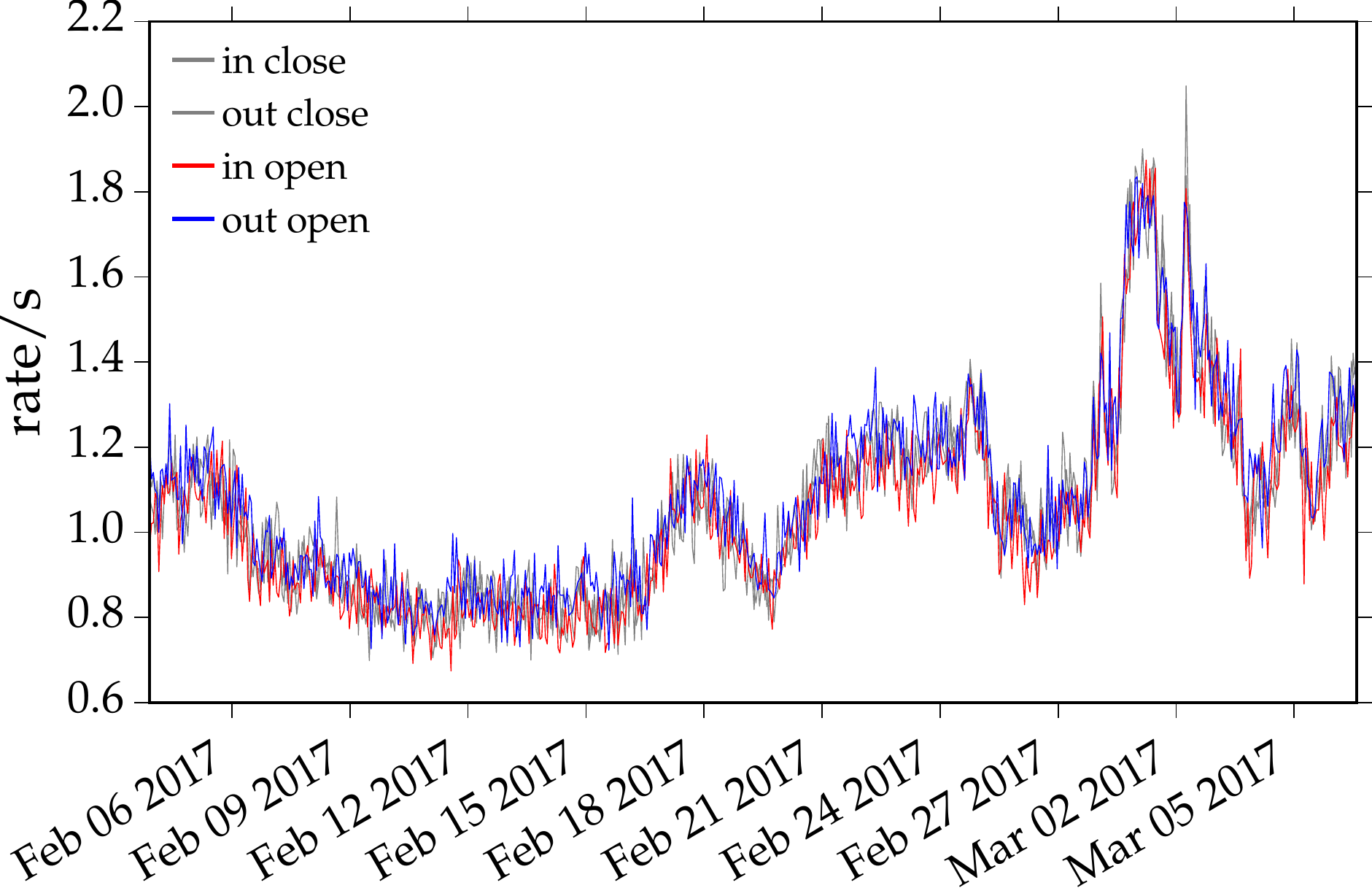}
\caption{Raw count-rate for individual 60\,s runs of the February
2017 measurement.}
\label{f:rate}
\end{figure}

\begin{figure}[t]
\centering
\includegraphics[height=0.4\textwidth]{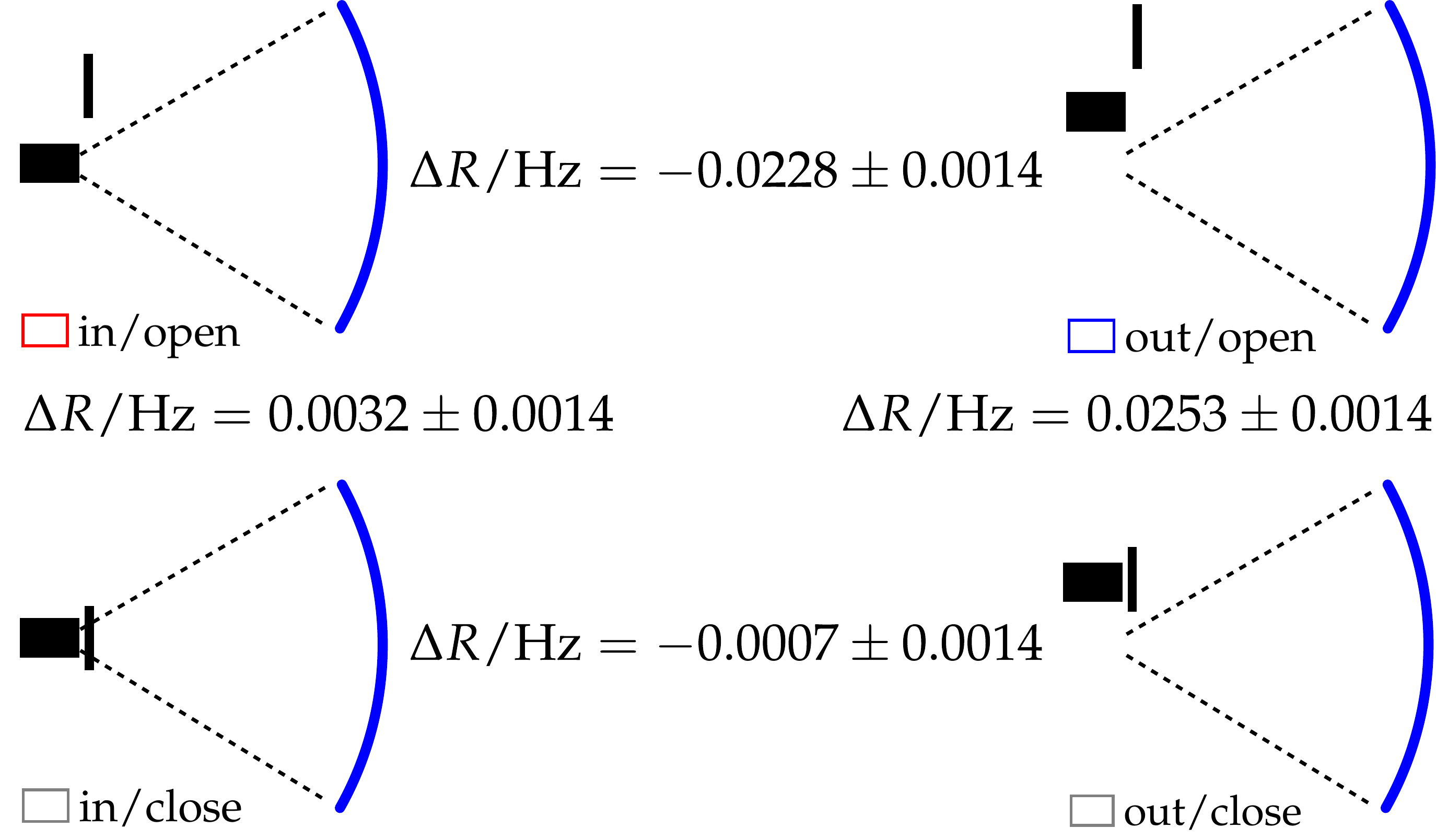}
\caption{Differences in the counting rate with the PMT in in/out position with open/close shutter.}
\label{f:measurement}
\end{figure}

\section{The \funk experiment}

In \funk, a spherical mirror is used for searching for
HP~\cite{Horns:2012jf,Jaeckel:2013sqa}.  When HP cross the metallic surface of
the mirror, faint electromagnetic waves are emitted almost perpendicularly to
the conducting surface.  In the case of a spherical mirror these photons are
focused on the radius point of the mirror.  Placing various detectors in the
radius point allows us to search for a photon signal in different wavelength
ranges, corresponding to different HP masses.  The DM nature of a possible
signal can be verified by observation of the expected small daily and seasonal
movements of the spot~\cite{Jaeckel:2013sqa,Dobrich:2014kda,Jaeckel:2015kea}.

\paragraph{Mirror.}

For the purposes of the \funk experiment we reused prototype mirror segments
that were originally produced for the Schmidt telescopes of the fluorescence
detector~\cite{Abraham:2009pm} of the Pierre Auger
Observatory~\cite{ThePierreAuger:2015rma}. At KIT, we assembled them inside of
a windowless air-conditioned experimental hall of ${\sim}20$\,m diameter with
2\,m thick concrete walls. The mirror represents a portion of the surface of a
sphere with a radius of 3.4\,m and total area of the mirror front surface is
14.56\,m$^2$ (see \fref{f:mirror}). For a more detailed description of the
mirror we refer to the Ref.~\cite{Veberic:2015yua}. The whole experimental area
with dimensions $495{\times}431{\times}435$\,cm$^3$ has been additionally
covered with a double-layer light shield (black plastic sheet over a cotton
curtain) and the ground was treated with a non-reflective black paint, as can
be seen in \fref{f:setup}.

To quantify the optical quality of the mirror we built a movable imaging
platform at the center of the mirror sphere. It consisted of a SMD LED as a
light source, placed very close to the center of the sphere, which projected an
image onto a frosted-glass screen. This image was then observed with a CCD
camera (see the three insets in \fref{f:focus}-left) and a movable linear stage
enabled us to obtain a cross-section of the converging light beams from
individual mirror segments. The mirror segments were adjusted individually to
minimize the size of the image. The \fref{f:focus}-left shows a scan through
the central region after alignment, with contour lines in steps of 10\%
indicating the amount of light for various distances away from the central
point. As can be seen in \fref{f:focus}, the spot that contains 90\% of all
light has a radius of ${\sim}2$\,mm. Note that the seasonal oscillation of the
HP spot due to the relative movement of the mirror in the DM frame is of the
same order~\cite{Jaeckel:2013sqa}.

\paragraph{Photomultiplier.}

After the preparation of the mirror we performed a test measurement with a
cooled PCO sensicam VGA CCD camera, the results of which have been presented in
Ref.~\cite{Veberic:2015yua}.  For the present measurement we chose a 29\,mm
diameter photomultiplier (PMT) ET 9107BQ~\cite{et-manual} with a blue-green
sensitive bialkali photocathode with the quantum efficiency extended into the
ultra-violet range (see \fref{f:mirror}-right).  The PMT has a high gain at
relatively low voltages, peak quantum efficiency of around 28\%, and excellent
single electron and pulse-height resolution, which makes it well-suited for
photon counting applications.  The active part of the photocathode has a
diameter of 25\,mm and is thus large enough to capture the spot of the
potential DM signal even in the presence of seasonal movements.

The PMT was placed into a Fact-50 enclosure (see \fref{f:setup}-left), which is
capable of lowering the temperature of the PMT thermoelectrically by up to
50\,K, relatively to the ambient.  Nevertheless, for the preliminary
measurement presented in this proceeding, the cooling capability was not yet
used, mostly because we wanted to avoid any unneccessary glass material in
front of the PMT which could produce Cherenkov light by cosmic-ray muons
passing through.

The PMT camera has been placed at the center of the spherical mirror (see
\fref{f:mirror}-left) and equipped with an additional Thorlabs motorized
flipper MFF101 carrying a shutter that can obscure the opening of the PMT to a
very high degree. The operating high-voltage setting was chosen to be 1050\,V,
close to a knee feature observed in a voltage scan of the PMT dark-current
rate~\cite{Dobrich:2015tpa}.

\paragraph{Data acquisition.}

Signals from the PMT were digitized with the PicoScope 6404D~\cite{picoscope}.
Examples of triggered traces can be seen in \fref{f:traces}, where a nice,
single-photon (SP) pulse can be observed on the left, while on the right we are
showing an example of a trace that contained several pulses in the
3.2\,$\upmu$s trigger window. Such traces were discarded since they clearly
belong to signals produced by cosmic-ray showers. Based on measurements of the
SP charge spectrum with an LED flasher setup (\fref{f:charge}-left), we also
applied a cut on the allowed range of observed charges, as seen in
\fref{f:charge}-right. From the SP measurements we also estimated that the
efficiency of such a cut on a SP signals is 75\%.

\section{Preliminary results}

\noindent
The resulting counts for the February 2017 run can be seen in \fref{f:rate}.
The data was taken in cycles of 60\,s in the two positions of the PMT camera,
with the sensitive area ``\emph{in}'' the center of the spherical mirror and
8\,cm ``\emph{out}'' of the center, and with the shutter ``\emph{open}'' or
``\emph{close}'' in both positions, thus resulting in four possible
combinations, as schematically shown in \fref{f:measurement}. The average rate
of the whole February run, after the quality selections, is $R=0.535$\,Hz and the
differences between the rates in the four different configurations are shown in
\fref{f:measurement}. 

\begin{figure}[t]
\centering
\includegraphics[width=0.8\textwidth]{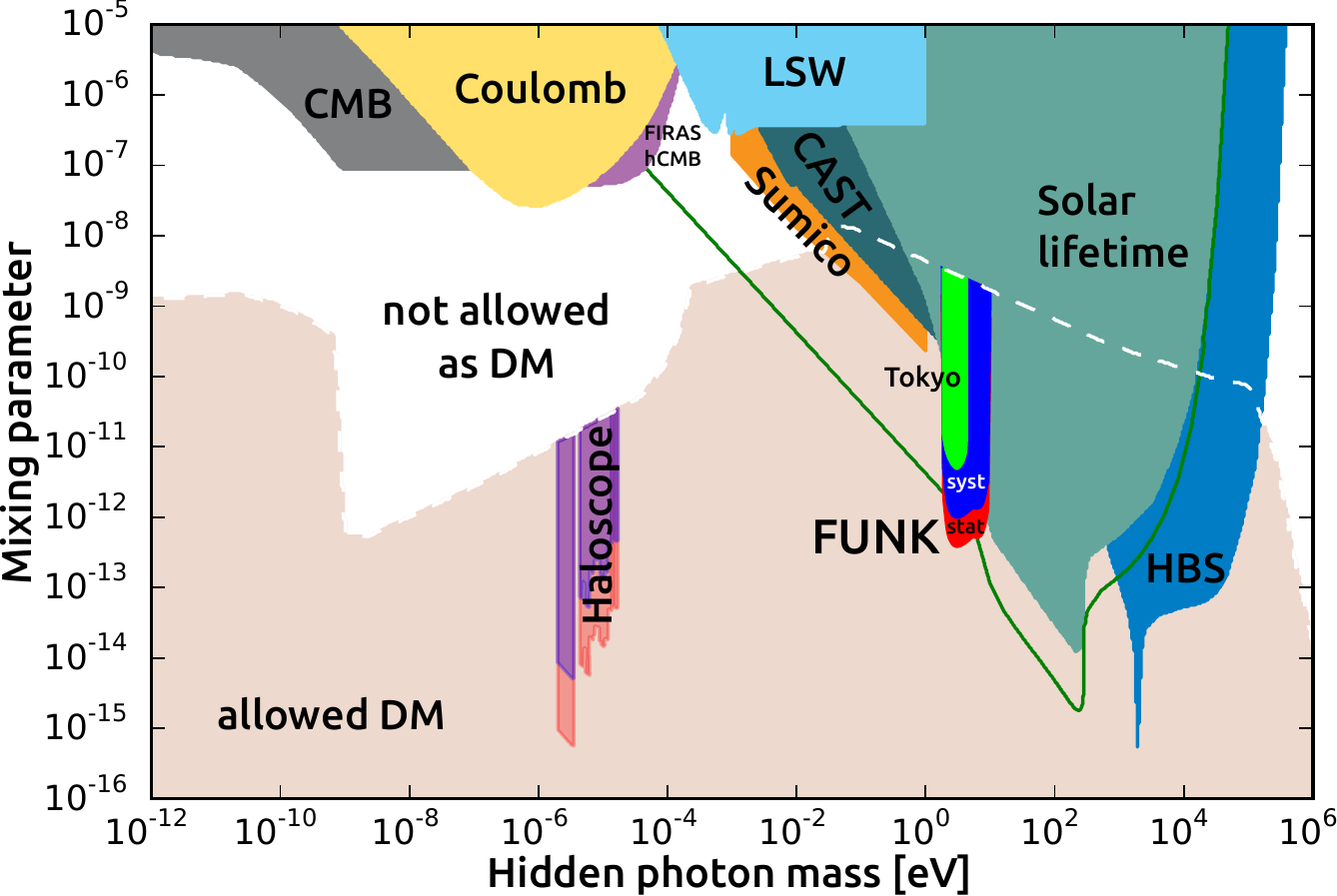}
\caption{This Figure is an adapted and updated version of the figure
in~\cite{Jaeckel:2010ni,Jaeckel:2013ija}, with the dark matter area
from~\cite{Arias:2012ly} and including new data and limits
from~\cite{An:2013vn,Vinyoles:2015aba}, in particular results from the Tokyo
dish experiment~\cite{Suzuki:2015sza,Suzuki:2015vka} (light green).  The
preliminary result for the \funk exclusion area in the mixing parameter vs.\
hidden photon mass parameter space is shown in blue (for the systematics-bound
result) and in red (for the statistics-bound result). Meaning of some other
labels: measurements on cosmic microwave background (CMB),
light-shining-through-wall experiments (LSW), horizontal-branch stars (HBS),
and the green line corresponds to improved solar
limits~\cite{An:2013yfc,Redondo:2013lna}; for summary and labels
see~\cite{Goodsell:2009xc}.}
\label{f:result}
\end{figure}

The difference of the count rates between ``\emph{open}'' and ``\emph{close}''
for the PMT being in the radius point is the measurement we are interested in.
With the PMT window open there are $\Delta R = 0.0032 \pm 0.0014$\,Hz more
counts registered than with the PMT closed. If we ignore any possible sys.{}
uncertainties, we would obtain the limit shown in \fref{f:result} as ``stat''
for \funk. For determining possible sys.{} uncertainties that might be related
to temperature changes or the limited accuracy of determining the measurement
time, we also compare the rates with the covered PMT in and out of the radius
point. The two count rates agree very well within the stat.{} uncertainty
($\Delta R = 0.0007 \pm 0.0014$\,Hz). Another cross check is the comparison of
the count rates for the open PMT in and out of the radius point. We found a
significant larger count rate if the PMT is open and outside of the radius
point, which is possibly related to the different imaging properties of the
setup in the two positions. Measurements are in progress to better understand
this observation. For the time being we treat this difference of rates ($\Delta
R \approx 0.025$\,Hz) as an upper limit of the overall sys.{} uncertainty of
our measurements and use this number to derive a preliminary upper
limit~\cite{Dobrich:2014kda} on the magnitude of the mixing parameter $\chi$ in
the sensitivity range of the PMT, see the \funk ``syst'' in \fref{f:result}.

\section{Conclusions and future plans}

\noindent
We performed a search for HP dark matter with a metallic spherical mirror of
$14$\,m$^2$ in the visible and near-UV wavelength range.  No signal was found.
Because the detailed analysis of the data is still ongoing we have reported
here only preliminary results with a maximally conservative estimate of
possible systematic uncertainties.

In the near future we are planning further measurements a fully cooled PMT, for
which we expect reduced number of background counts. In addition we intend to
extend the search for possible HP-DM  into the MHz~\cite{Aab:2014esa},
GHz~\cite{Smida:2014sia}, and THz range.

\paragraph{Acknowledgments.}

We gratefully acknowledge partial support from the Helmholtz Alliance for
Astroparticle physics (HAP), funded by the Initiative and Networking Fund of
the Helmholtz Association.

\end{document}